\documentclass[preprint,preprintnumbers,amsmath,amssymb]{revtex4}
\usepackage{graphicx}
\usepackage{dcolumn}
\usepackage{bm}
\def\Dagcomp{^{\vphantom{\dagger}}}

\begin{document}

\title{Quantum transport anomalies in DNA containing mispairs}
\author{Xue-Feng Wang}
\email{xf_wang1969@yahoo.com}
\affiliation{Department of physics,
Soochow University, 1 Shizi Street, Suzhou, 215006 China}
\author{Tapash Chakraborty and J. Berashevich}
\affiliation{Department of Physics and Astronomy, The University
of Manitoba, Winnipeg, Canada, R3T 2N2}

\begin{abstract}
The effect of mispair on charge transport in a DNA of sequence (GC)(TA)$_N$(GC)$_3$ connected to platinum electrodes is studied using the tight-binding model. With parameters derived from {\it ab initio} density functional result, we calculate the current versus bias voltage for DNA with and without mispair and for different numbers of (TA) basepairs $N$ between the single and triple (GC) basepairs. The current decays exponentially with $N$ under low bias but reaches a minimum under high bias when a multichannel transport mechanism is established. A (GA) mispair substituting a (TA) basepair near the middle of the (TA)$_N$ sequence usually enhances the current by one order due to its low ionization energy but may decrease the current significantly when an established multichannel mechanism is broken.
\end{abstract}
\maketitle

\section{introduction}

Longitudinal charge transport along DNA has been the subject of extensive study in the last decade. \cite{schu,chak,gene}
Charge transport occurs in the oxidative and reductive DNA damage or repair processes and can happen in the long distance range. \cite{brow,loft,schu} Study of transport properties may lead to a better understanding of the fundamental driving processes in biological evolution. Furthermore, the charge transport process might have been used naturally for basepair mismatch detection during the DNA repairing process. It is already known that, due to chemical reaction and radiative ionization, mispairs or gene mutations happen quite often in the cell. Fortunately, almost all of the mispairs can be detected and repaired during the replication process to keep the material genetically stable. However, some of the mutations may escape from the detecting and repairing processes and result in various genetic diseases including cancer. A recent study indicates a negative correlation between the cancer risk and sensitivity of charge transport property of the gene to a mutation. \cite{cancer}
Understanding how mispairs modify the electric properties of DNA then becomes very important \cite{scha,okad,edir} and, together with the usage of other properties, \cite{apal,bera} may improve mutation detecting techniques. \cite{fixe,marr,tian,zhan}  In addition, thanks to its perfect self-assembling and self-recognition properties found in nature, DNA is also expected to be a potentially functional material for molecular devices. In this case mispairs may be used to obtain unique functions of the devices.

The charge transport through a DNA sequence can be measured by chemical or physical methods. \cite{schu,chak,gene} In one of the typical chemical experiments, Giese et al. used a DNA of sequence (GC)(TA)$_N$(GC)$_3$. \cite{gies} They measured the charge transfer rate from the (GC) basepair to the (GC)$_3$ triple basepair for different number $N$ of (TA) basepairs, and found a crossover from a rapid decay of the charge transfer rate vs $N$ to an almost zero decay around $N=3$.
As an alternative to other explanations, \cite{berl,bixo,jort,reng,cram,bask} we have proposed this as a crossover from one dominant channel transport to a multichannel transport. \cite{wang2} An example of physical experiments is the one performed by Porath et al.\cite{pora1} where a DNA sequence (GC)$_m$ is located between two platinum electrodes and the current versus voltage is directly measured.  This result has also been simulated by simple tight-binding models. \cite{li,cuni,wang1} It is known that G.A mispair in various conformations \cite{bera} is the most stable mispair and often present in the DNA. \cite{brow} The magnetic properties of DNA was studied earlier and found to be significantly influenced by the presence of the G.A mispair. \cite{apal} In this paper, we will study the effect of mispairs, such as G(anti)$\cdot$A(anti) indicated in the following as (GaAa), and G(anti)$\cdot$A(syn) indicated as (GaAs), \cite{leon} on charge transport when a Watson-Crick (TA) basepair is replaced by a mispair in the DNA sequence (GC)(TA)$_N$(GC)$_3$ connected to platinum electrodes.

\section{method}

\begin{figure}
\begin{center}
\begin{picture}(330,235)
\put(-10,90)
{\includegraphics{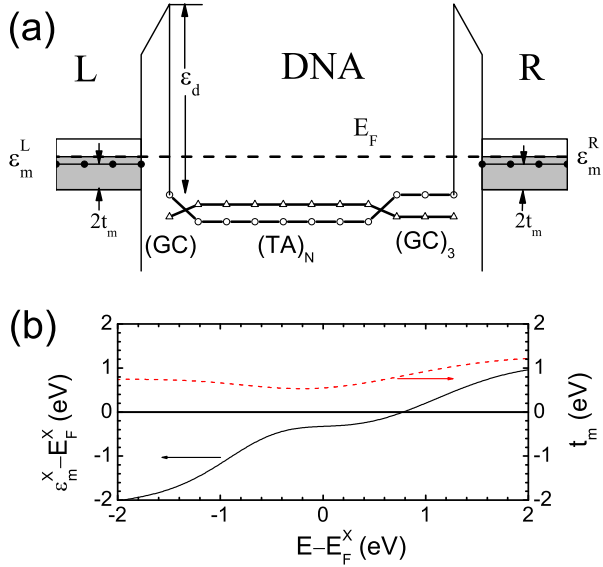}}
\end{picture}
\vspace{-1.0cm}
\protect\caption{(a) Schematic illustration of the equilibrium energy band across the system with a DNA of sequence (GC)(TA)$_N$(GC)$_3$ connect to two platinum electrodes. (b) The energy dependence of $\epsilon_m$ (solid curve) and $t_m$ (dashed) for electrons in the platinum electrode $X$, with $X=L$ or $R$ for the left or right electrode.}
\label{fig:fig1}
\end{center}
\end{figure}

We consider a $p$-type semiconductor DNA duplex chain of basepairs
connected to a circuit via two platinum electrodes suitable for experimental realization \cite{pora1}.
Each platinum electrode is modeled as a semi-infinite one-dimensional (1D) electrode \cite{wang1} connected to the G base at one end of the first strand as illustrated in \ref{fig:fig1}(a).
The tight-binding Hamiltonian of the system reads

\[H =2\sum_{n=-\infty}^\infty
[\varepsilon _n c_n^\dag c_n
-t_{n,n+1} (c_n^\dag c_{n+1}+c_{n+1}^\dag c_n)]
\]
\[+2\sum_{n=1}^N u_n d_n^\dag d_n
-2\sum_{n=1}^{N-1} h_{n,n+1}(d_n^\dag d_{n+1}+d_{n+1}^\dag d_n)
\]
\begin{equation}
-2\sum_{n=1}^N
\lambda_n (c_n^\dag d_n+d_n^\dag c_n).
\label{eq:hami}
\end{equation}

Here $c_n^\dag$ ($d_n^\dag$) is the creation operator of holes
in the first (second) strand on site $n$ of the DNA chain (for
$1\leq n \leq N$), the left electrodes ($n\leq 0$), and the right
electrodes ($n\geq N+1$). The on-site energy of site $n$ in the
first (second) strand is denoted by $\varepsilon_n$ ($u_n$), which
is equal to the highest occupied molecular orbit (HOMO) energy of
the base on this site in the DNA chain and the center of conduction
band in the electrodes. The coupling parameter of the first (second)
strand $t_{n,n+1}$ ($h_{n,n+1}$) is equal to the intra-strand
coupling parameter between neighboring sites $n$ and $n+1$
of the DNA for $1\leq n \leq N-1$, one-fourth of the conduction band-width
in the electrodes $t_m$ for $n \leq -1$ and $n\geq N+1$, and the
coupling strength between the electrodes and the DNA strands
for $n=0$ and $n=N$. The inter-strand coupling between sites in the
same basepair is described by $\lambda_n$. The factor
2 multiplied to each sum in Eq.\,(\ref{eq:hami}) arises from
the spin degeneracy.

In transport experiments,\cite{pora1} a high bias voltage can be applied to drive the system
far from equilibrium and holes in wide energy range may contribute to the current.
Since the carriers usually come from various energy bands and the profile of band distribution is energy dependent,
the effective parameters $\varepsilon_{m}$ and $t_{m}$, which are averages
over the profiles, are then energy dependent. We assume that the parameters
for the 1D tight-binding model have a similar dependence on energy
as in bulk platinum \cite{roge} and the dependence is extracted from its 3D band structure.
Near the Fermi energy, there are six bands located approximately at
$-5.8$, $-4.7$, $-3.7$, $-2.2$, $-0.2$, and $2.0 $eV above the Fermi energy with band
width $1.9$, $1.3$, $1.5$, $3.1$, $1.4$, and $6.0$ eV respectively. Using Lorentzian
broadening, we can mimic the bulk DOS and extract the parameters
$\varepsilon_{m}$ and $t_{m}$ as shown in \ref{fig:fig1}(b).
The parameters are then scaled to match the known values at the Fermi energy as was done in Ref.\cite{wang1}. For electrons at the Fermi energy, the on-site energy is
$\epsilon^0_m=-0.33$ eV with a coupling parameter $t^0_m=0.55$ eV.
As estimated from the experimental data \cite{li,wang1} the equilibrium Fermi energy is 1.73 eV higher than the
HOMO on-site energy of the G base when the (G$\cdot$C) basepair makes contact with the platinum electrodes. Here we assume that the first DNA strand is coupled to the electrodes with a contact parameter of $t_{dm}=0.1$ eV while the second strand does not contact the electrodes directly. Note that our main result is not sensitive to the choice of the electrodes and the contact parameters.

\begin{figure}
\begin{center}
\begin{picture}(330,120)
\put(-0,0)
{\includegraphics{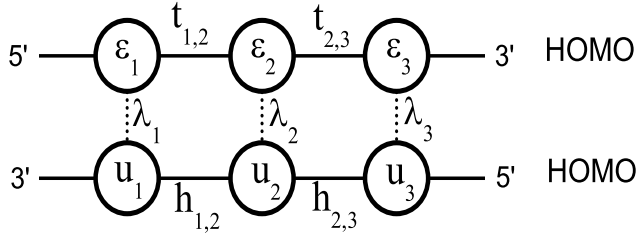}}
\end{picture}
\vspace{-1.0cm}
\protect\caption{Illustration of a three-basepair DNA used in the ADF program to obtain the tight-binding parameters.
In the first (second) strand, the on-site energy of a HOMO orbital is $\varepsilon_n$ ($u_n$) and the intrastrand coupling parameter between neighboring sites is $t_{n,n+1}$ ($h_{n,n+1}$) with $n$ the base index.
The interstrand coupling parameter is denoted by $\lambda_n$.}
\label{fig:fig2}
\end{center}
\end{figure}

The tight-binding parameters of DNA are estimated based on the HOMO energies of isolated nucleobases and the charge transfer integral between the HOMO orbitals calculated by the {\it ab initio}
density functional method integrated in the ADF (Amsterdam Density Functional) program. \cite{adf,apal,bera} The on-site energies for bases G, C, T, and A are $-9.40$, $-10.27$, $-10.46$, and $-9.79$, respectively.
The hopping coupling parameters are listed in Table \ref{tab:table2}.

\begin{table}
\caption{\label{tab:table2}
The values of the intra- and interstrand tight-binding parameters (in eV) for
different DNA sequences, where the middle pair (X$_1$$\cdot$X$_2$)
can be G$\cdot$C and T$\cdot$A basepairs or G(anti)$\cdot$A(anti) and
G(anti)$\cdot$A(syn) mispairs. \cite{apal}}
\begin{tabular}{c|ccccc}
\multicolumn{6}{c} {5'-G$-$X$_1$$-$G-3'}\\
\multicolumn{6}{c} {3'-C$-$X$_2$$-$C-5'}\\
\hline
(X$_1$$\cdot$X$_2$) &$t\Dagcomp_{\rm 12}$ & $t\Dagcomp_{\rm 23}$
& $\lambda\Dagcomp_{\rm 2}$ & $h\Dagcomp_{\rm 12}$
& $h\Dagcomp_{\rm 23}$ \\
\hline \\
G$\cdot$C& 0.133 & 0.133 & 0.028 & 0.14 & 0.14 \\
T$\cdot$A& 0.164 & 0.400 & 0.070 & 0.154 & 0.099\\
\hline
\multicolumn{6}{c} {}\\
\multicolumn{6}{c} {5'-T$-$X$_1$$-$T-3'}\\
\multicolumn{6}{c} {3'-A$-$X$_2$$-$A-5'}\\
\hline
(X$_1$$\cdot$X$_2$) &$t\Dagcomp_{\rm 12}$ & $t\Dagcomp_{\rm 23}$
& $\lambda\Dagcomp_{\rm 2}$ & $h\Dagcomp_{\rm 12}$
& $h\Dagcomp_{\rm 23}$ \\
\hline \\
T$\cdot$A & 0.330 & 0.330 & 0.070 & 0.011 & 0.011\\
Ga$\cdot$Aa& 0.400 & 0.164 & 0.029 & 0.055 & 0.002\\
Ga$\cdot$As& 0.250 & 0.167 & 0.057 & 0.207 & 0.027\\
\end{tabular}
\end{table}

The current $I$ when a voltage bias $V$ is applied over the two platinum electrodes is then evaluated by the transfer matrix method
\cite{wang2,chak,yan,marc}. For an open system, the secular equation
is expressed as a group of equations of the form
\begin{eqnarray*}
t_{n-1,n}\Psi^+_{n-1}+(\varepsilon_n-E)\Psi^+_n
+\lambda_n \Psi^-_n+t_{n,n+1}\Psi^+_{n+1}=0\\
h_{n-1,n} \Psi^-_{n-1}+(u_n-E) \Psi^-_n
+\lambda_n \Psi^+_n+h_{n,n+1} \Psi^-_{n+1}=0
\end{eqnarray*}
with $\Psi^+_n$ ($ \Psi^-_n$) the wave function of the first (second)
strand on site $n$. The wave functions of the sites $n+1$ and $n$ are
related to those of the sites $n$ and $n-1$ by a transfer matrix $\hat{M}$,
\begin{equation}
\left( \begin{array}{c}
\Psi^+_{n+1}\\
\Psi^-_{n+1}\\
\Psi^+_n\\
 \Psi^-_n
\end{array}\right) =\hat{M}
\left(
\begin{array}{c}
\Psi^+_n\\
 \Psi^-_n\\
\Psi^+_{n-1}\\
 \Psi^-_{n-1}
\end{array}\right),\,\,\,
\label{eq:matrix}
\end{equation}
with
\begin{equation*}
\hat{M}=
\left[ \begin{array}{cccc}
\frac{(E-\varepsilon_n)}{t_{n,n+1}}& \frac{-\lambda_n}{t_{n,n+1}}&
-\frac{t_{n-1,n}}{t_{n,n+1}}&0\\
\frac{-\lambda_n}{h_{n,n+1}}&\frac{(E-\varepsilon_n)}{h_{n,n+1}}
&0&-\frac{h_{n-1,n}}{h_{n,n+1}}\\
1&0&0&0\\
0&1&0&0
\end{array}\right].
\end{equation*}
The transmission is then calculated by assuming the plane waves
propagating in the electrodes for the holes
$\Psi_n=A e^{ik_Lna}+B e^{-ik_Lna}$ for $n\leq 0$ and
$\Psi_n=C e^{ik_Lna}$ for $n\geq N+1$ in the left and right electrodes,
respectively.
Expressing the output wave amplitude $C$ in
terms of the input wave amplitude $A$ and the transmission,
$$T(E)=\frac{|C|^2\sin(k_R a)}{|A|^2\sin(k_L a)}.$$
The distance between two neighboring bases along any DNA strand is $a=3.4$ \AA. The net current primarily comes
from the hole transmission between the electrodes' Fermi
energies and is calculated as \cite{datt}
$$I=\frac{e^2}{h}\sum_\sigma \int_{-\infty}^\infty
dE\, T^\sigma (E)[f^L(E)-f^R(E)].$$ Here the Fermi function is
$f^X(E)=1/\exp[(E-E^X_F)/k_B T]$ with $X=L$ or $R$ and the room temperature $T=300$ K.
When a bias voltage $V$ is applied between the two electrodes, the left (right) Fermi energy
is assumed $E^L_F=V/2$ ($E^R_F=-V/2$).

\section{results and discussions}

\begin{figure}
\begin{center}
\begin{picture}(330,170)
\put(-10,145)
{\includegraphics{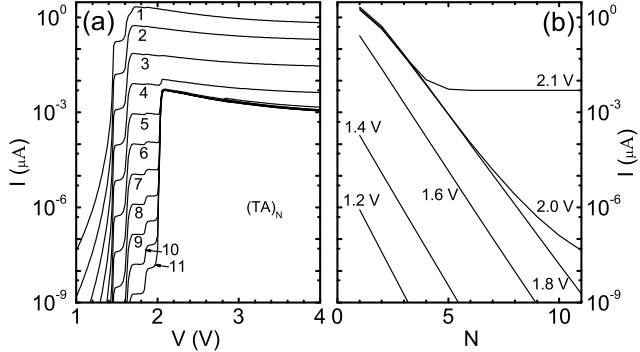}}
\end{picture}
\vspace{-1.0cm}
\protect\caption{(a) Current $I$ versus bias voltage $V$ of the DNA sequence
 (GC)(TA)$_N$(GC)$_3$. $N=1-11$ for curves from the top. (b) Current I versus (TA) basepair number $N$ at fixed bias voltage $V$. The value of $V$ is indicated beside each curve.}
\label{fig:fig3}
\end{center}
\end{figure}

The results for the current ($I$) versus voltage ($V$) for DNA of sequence
(GC)(TA)$_N$(GC)$_3$ with
$N=1,2,...,11$ are shown in \ref{fig:fig3}(a). Each curve except for $N=1$ has steps at voltages 1.45, 1.64, 1.85, and 2.02 eV indicating that the transport channels are mainly formed at four threshold voltages related to the four kinds of bases. In the $N=1$ curve, the second step is split into two at $V=1.62$ and 1.7 V and there is no step at 2.02 V. In a log scale, the two lower-energy steps have almost the same height for all the curves while the height of the two higher-energy steps increases with the number of (TA) base-pairs. At a bias voltage lower than the first threshold, the DNA works as an energy barrier for electron transport since the Fermi energy of both electrodes is located between the HOMO and LUMO energy. At a bias near the first and the second threshold ($1.45<1.8$ eV), HOMO channels of the (GC) basepairs become available for charge transport but the (TA) basepairs behave as energy barriers for transport. At a bias higher than the third threshold ($V>1.85$ eV), transport channels of (TA) basepair also participate in the charge transport across the DNA molecule. Along the curve $N=1$, we can hardly see the third and the fourth steps. The addition of (TA) basepairs can establish a network of bases due to the interstrand coupling and introduces additional transport channels, as was reported in Ref.\cite{wang2}. Consequently, we observe the height increase of the third and fourth steps in the log scale. The current enhancement due to additional channels may compete with the exponential current decay with the length of the molecule. This results in a current minimum for $N>5$, as clearly shown in \ref{fig:fig3}(b) where the current is plotted versus $N$ at various $V$. At a bias less than 1.8V, the current decays exponentially with $N$ with almost the same exponent. For a bias higher than 1.8V, the exponent decreases with $N$ until it is almost zero for $N>5$ under the bias $V= 2.1$V or higher. This crossover from a rapid to almost zero decay of the charge transfer versus the (TA) basepair numbers has been observed in Ref.\,\cite{gies} with the chemical method and can also be observed in physical experiments as described in Ref.\,\cite{pora1}. Different from our previous simplified model \cite{wang2} where uniform parameters and virtual electrodes are assumed, here we employ a more
realistic model with the tight-binding parameters of DNA and electrode extracted
from ab initio calculations. In addition, a variable bias voltage is applied between the two
electrodes to obtain the I-V curve. Note that the role of diagonal interstrand hopping is relevant to the electron transport in DNA and its inclusion in the calculation might shift the I-V curve but not the conclusion. \cite{well,wang3} 

\begin{figure}
\begin{center}
\begin{picture}(330,210)
\put(-16,130)
{\includegraphics{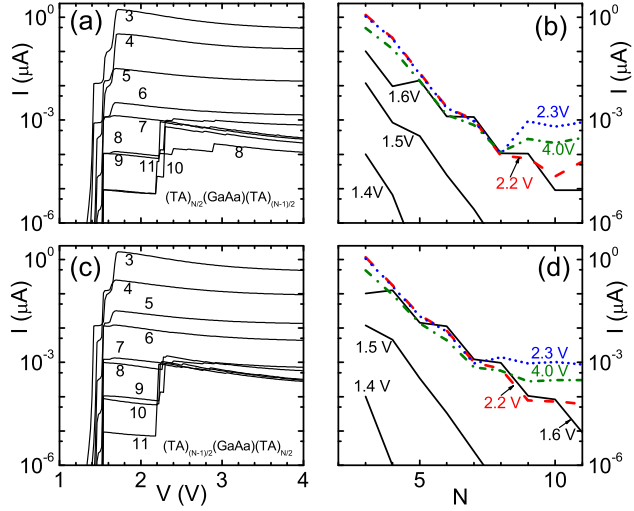}}
\end{picture}
\vspace{-1.0cm}
\protect\caption{(a) The $I-V$ curves of DNA sequence
 (GC)(TA)$_{N/2}$(GaAa)(TA)$_{(N-1)/2}$(GC)$_3$ for $N=3,4,...,11$. Here $N/2$ and $(N-1)/2$ take only the interger part of the value. (b) Current I versus the number $N$ at several selected bias voltage $V$ indicated by the values beside each curve. The $I-V$ curves and $I-N$ curves  for DNA sequence (GC)(TA)$_{(N-1)/2}$(GaAa)(TA)$_{N/2}$(GC)$_3$ are plotted in (c) and (d).}
\label{fig:fig4}
\end{center}
\end{figure}

To estimate the effect of mispairs (GaAa) and (GaAs) on the charge transport, we replace one of the (TA) basepairs near the middle of the (TA)$_N$ sequence by a mispair and calculate the corresponding $I-V$ curve. In \ref{fig:fig4}(a) and (b), the result for the [$(N+2)/2$]th (TA) basepair replaced by a (GaAa) mispair is shown and in (c) and (d) the [$(N+1)/2$]th basepair is replaced by a mispair. Here [$R$] means extracting the integer part of a real number $R$.
Note that the curves of odd $N$ are the same in \ref{fig:fig4}(a) and (c), corresponding to equal number of (TA) basepairs to the left and right of the (GaAa) mispair. For even $N$, there are one more (TA) to the left of the mispair in (a) and one less in (b).
With the replacement of the mispair, the current is greatly enhanced after the first threshold voltage, indicating that the G base in the mispair works as a tunneling bridge. However, for $N > 5$, the current decreases with the mispair at a bias higher than the fourth threshold voltage (2 eV) as shown in \ref{fig:fig4}(a) and (c). This happens because the mispair destroys the resonant transport network formed by the periodic (TA) basepairs series, i.e. the mispair works as an impurity. A weak $N$ dependent current appears only for higher $N$ and at a lower current value, as shown in \ref{fig:fig4}(b) and (d) when both the (TA) sequences besides the mispair form resonant transport network.

In the presentce of a mispair, some steps shift and extra steps appear along the $I-V$ curves. For $N=3$ or a DNA of sequence (GC)(TA)(GaAa)(TA)(GC)$_3$, the simplest case with a mispair, three steps at $V=1.41$, 1.55, and 1.68 V appear in the $I-V$ curve and the current is enhanced by more than one order at high voltage with the substitution by a mispair. For $N\ge 5$, however, the first three steps shift to $V=1.45$, $1.53$, and $1.64$ V with the last one decaying over $N$. Compared to the case without a mispair, one extra step appears at $V=1.53$ V. In the bias range 1.45 V $<V<$ 1.53 V, the curves for $N>4$ are almost equally separated in the vertical log scale, indicating an exponentially decaying current with $N$ as also shown in \ref{fig:fig4}(b) and (d). Furthermore, the mispair location also affects significantly the current. The $N=4$ curve has steps at the same position as for $N=3$ in \ref{fig:fig4}(a) but as the $N\ge 5$ curves in \ref{fig:fig4}(b). In the range 1.58 V $<V<$ 2.2V, the even $N$ curves are near the curve of $N+1$ in (a) but near the curve $N-1$ in (b). The change is also represented by shifted steps along the $V=1.6$ V curves when comparing \ref{fig:fig4}(b) to (d). This observation suggests that the current in this bias range is mainly determined by the length of the (TA)$_n$ sequence between the left single (GC) basepair and the (GaAa) mispair. Under stronger bias, the current at first decays exponentially with $N$ and then fluctuates near a value slightly below 1 nA. This long-DNA current limit is about five times smaller than that observed in the system without a mispair.

\begin{figure}
\begin{center}
\begin{picture}(330,210)
\put(-16,130)
{\includegraphics{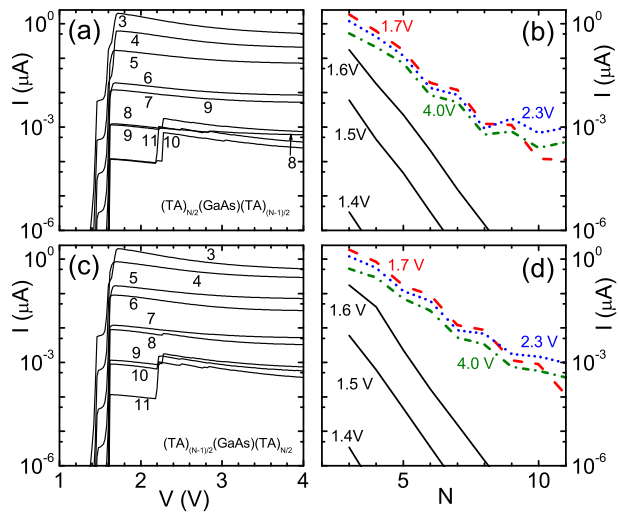}}
\end{picture}
\vspace{-1.0cm}
\protect\caption{The same as in \ref{fig:fig4} is plotted for basepair mismatch (GaAs).}
\label{fig:fig5}
\end{center}
\end{figure}

When the (GaAa) mispair is replaced by a (GaAs) mispair, the $I-V$ curves show fewer steps as illustrated in Fig.\,\ref{fig:fig5}. For $N=3$ there are three steps at $V=1.45$, $1.58$, and $1.68$V while for $N\ge 5$ there are only two steps at $V=1.45$ and $1.61$V below bias 2V. Similar to the case of (GaAa) mispair shown in \ref{fig:fig4}, the $N=4$ curve also has steps at the position of the $N=3$ curve in (a) but at the positions of the $N\ge 5$ curves in (b). In addition, the current is mainly limited by the number of (TA) basepairs between the left (GC) basepair and the (GaAs) mispair in the range 1.6V$ < V < 2.2$V. For $V>2.2$V, no step in curves of $N \le 7$ suggests again that one transport channel dominates in the short (TA)$_n$ DNA sequence. For large $N$, a series of steps appear in the $I-V$ curves and the current do not decrease monotonically with $N$, indicating the enhancement of current due to the increase of transport channel number with $N$ can compensate the current decay with the length of each channel.  The current decays exponentially at $V<1.6$ V but decays with steps in the range $1.6<V<2.2$ V. At $V>2.2$ V the current decays in short DNA and then fluctuates when multichannel tunneling mechanism dominates in the long (TA)$_n$ sequence.
Overall a (GaAs) mispair substitution changes less the current than that of a (GaAa) mispair especially in high bias since the intrastrand coupling parameter of a (GaAs) mispair is closer to that of a (TA) basepair.

\section{summary}

In summary, we have studied charge transport through DNA
connected to two platinum electrodes and contains mispairs within a realistic tight-binding scheme. The energy dependent tight-binding parameters for the electrodes are obtained by fitting the density of states near the Fermi energy of the material. The parameters of DNA are derived from the {\it ab initio} density functional calculation of the coupling between HOMO states in neighbor bases. When a (TA) basepair in the (GC)(TA)$_N$(GC)$_3$ sequence is replaced by (GaAa) or (GaAs) mispairs, the current is usually enhanced due to the lower ionization energy of the mispairs. In DNA with a long (TA)$_N$ sequence, multichannel tunneling mechanism set a minimal current at a high bias, similar to a previous experimental observation. The substitution of the mispair in this case, however, will break the multichannel tunneling mechanism and decrease the current significantly.

\section{acknowledgements}

X.F.W. acknowledges support from the startup fund in Soochow University and NSFC in China. The work was supported in part by the Canada Research Chairs Program (T.C.).

\end{document}